\documentclass[aps,twocolumn,pra,showpacs,superscriptaddress]{revtex4}
\usepackage{graphicx}
\usepackage{amsmath}
\begin{document}
\draft
\title{One-mirror Fabry-Perot and one-slit Young interferometry}

\author{Daniel Rohrlich, Yakov Neiman, Yonathan Japha, Ron Folman}

\address{Department of Physics, Ben-Gurion University \\
Beer-Sheva 84105 Israel}

\date{\today}


\begin{abstract}
We describe a new and distinctive interferometry in which a probe particle
scatters off a {\it superposition} of locations of a {\it single} free
target particle.  In one dimension, probe particles incident on superposed
locations of a single ``mirror" can interfere as if in a Fabry-Perot
interferometer; in two dimensions, probe particles scattering off
superposed locations of a single ``slit" can interfere as if in a two-slit
Young interferometer.  The condition for interference is {\it loss} of
orthogonality of the target states and reduces, in simple examples, to
{\it transfer} of orthogonality from target to probe states. We analyze
experimental parameters and conditions necessary for interference to be
observed.
\end{abstract}

\maketitle

     The two-slit interference experiment contains a mystery of quantum
theory, and Feynman even stated that ``it contains the {\it only}
mystery" \cite{feynman}.  Whether or not we accept Feynman's
statement, we can easily accept the importance of the two-slit
experiment and its generalizations to quantum theory.  Let us
consider a particularly ``quantum" generalization of the two-slit
experiment:  instead of two slits for the interfering quanta, the
experiment contains a single free ``quantum slit" (or a single
Fabry-Perot (FP) mirror) in a superposition of two locations.  Can
scattering from such a superposition show quantum interference?
Cohen-Tannoudji et al. \cite{c} answered this question \cite{q} in
the negative, with an assumption that is justified in specific
experimental settings. But the advent of new experimental
settings, such as one- and two-dimensional potentials on the Atom
Chip \cite{ron} and highly controlled atom optics, lead us to
reconsider the question.  Here we derive a general condition for
quanta impinging on a superposition of target locations to
interfere, and describe the experimental conditions for this
distinctive quantum interference to be observed.

     To specify the experimental setting, we replace the FP
mirrors, or the slits in the Young double-slit experiment, with a
single quantum target:  a scattering center, i.e. an ultracold
atom, in a superposition of orthogonal position states. Both the
probe and the target are free. We confine the target to move in
one dimension (i.e. in a tight atomic guide). In the first example
below, both the probe and target are one-dimensional, and the
superposed locations of the target form a one-dimensional,
one-mirror FP interferometer. In the second example, the probe
moves in a plane containing the target axis and scatters in two
dimensions off the superposed locations of the target, which form
a double-slit interferometer made of a single slit.

     Let the initial target wave function $\varphi_\alpha ({\bf X})$ be a
superposition of wave packets separated by a distance $d$,
\begin{equation}
\varphi_\alpha ({\bf X}) = {1\over\sqrt{2}}  \left[ \varphi_L ({\bf X})
+e^{i\alpha} \varphi_R ({\bf X}) \right] ~~~,\label{defin}
\end{equation}
where we take $\varphi_R ({\bf X}) =\varphi_L ({\bf X-d}) =e^{{\bf \nabla
\cdot d}} \varphi_L ({\bf X})$ for convenience.  The wave packets have
support in regions smaller than $d/2$. Such a wave function may be
engineered by growing a barrier in the middle of a harmonic oscillator
trap, so as to form a double well in the free dimension of the target, and
then by quickly shutting off this trap.

     How should quantum interference show up in scattering from a
superposition of locations?  An operational definition is
essential. No local measurement on $\varphi_L({\bf X})$ or
$\varphi_R ({\bf X})$ alone can yield $\alpha$, the relative phase
of the wave packets; no probe particle interacting with a target
at {\it one} of its locations, but not {\it both}, can provide any
information about $\alpha$.  Hence, dependence on $\alpha$ in the
final state of the probe is a sure signal of interference between
paths of the probe scattering from the two target locations.
Since $\varphi_\alpha ({\bf X})=[1 +e^{i\alpha +{\bf \nabla \cdot
d}}] \varphi_L ({\bf X}) /\sqrt{2}$, the Fourier transform
${\tilde \varphi}_\alpha ({\bf P})$ of $\varphi_\alpha ({\bf X})$
is
\begin{equation}
{\tilde \varphi}_\alpha ({\bf P}) =
[1+ e^{i\alpha +i{\bf P\cdot d}/\hbar}]
{\tilde \varphi}_L ({\bf P}) /\sqrt{2}
~~~,
\label{ft}
\end{equation}
and shows peaks in ${\bf P\cdot d}/d$ separated by $h/d$.  A change in
$\alpha$ shifts the peaks, i.e. changes the {\it modular} momentum
\cite{ar} defined as ${\bf P\cdot d}/d$ modulo $h/d$.  We will see how
$\alpha$ can show up in the final momentum distribution of the probe
particles.

     In the case that a probe and target have initial momenta ${\bf
p}^{in}$ and ${\bf P}^{in}$, respectively, the initial overall state
$\vert \Psi_{in}\rangle$ of the probe and target is
\begin{equation}
\vert\Psi_{in}\rangle =\vert {\bf p}^{in} \rangle \otimes\vert
\varphi_\alpha \rangle  =  \vert {\bf p}^{in} \rangle \otimes \int d^3
{\bf P}^{in}~ {\tilde \varphi}_\alpha ({\bf P}^{in})\vert {\bf
P}^{in}\rangle~,  \label{instate}
\end{equation}
In Eq.\ (\ref{instate}) and below, the first ket in any tensor product
refers to the probe and the second ket refers to the target. The state
$\vert {\bf p}^{in}\rangle\otimes\vert {\bf P}^{in}\rangle$ can scatter to
a state $\vert {\bf p}^{fin}\rangle\otimes \vert {\bf P}^{in}+{\bf p}^{in}
-{\bf p}^{fin}\rangle$.  We let $S({\bf p}^{in} , {\bf p}^{fin}; {\bf
P}^{in})$ denote the amplitude of the transition.  Then the overall final
state $\vert \Psi_{fin}\rangle$ is
\begin{eqnarray}
\int d^3{\bf p}^{fin} \int d^3 {\bf P}^{in} &{}& {\tilde \varphi}_\alpha
({\bf P}^{in}) S({\bf p}^{in} , {\bf p}^{fin}; {\bf P}^{in})
~\vert {\bf p}^{fin}\rangle\otimes \cr
&{}&\vert {\bf P}^{in}+{\bf p}^{in} -{\bf p}^{fin}\rangle~~~.
\label{finstate}
\end{eqnarray}
Cohen-Tannoudji et al. \cite{c} considered the limit in which $S({\bf
p}^{in} , {\bf p}^{fin}; {\bf P}^{in})$ is independent of ${\bf P}^{in}$,
a limit appropriate to photons scattering off a heavy atom.  With this
assumption, the overall final state $\vert \Psi_{fin}\rangle$ reduces to
\begin{equation}
\vert \Psi_{fin} \rangle = \int d^3 {\bf X} ~~\varphi_\alpha ({\bf X})
~~\vert \chi ({\bf X} ) \rangle\otimes \vert {\bf X} \rangle~~~,
\end{equation}
where $\vert\chi ({\bf X} )\rangle$ is a probe (photon) state that
depends on the location of the target.  They showed that there can be
no interference between photon states entangled with the two locations of
the target, because the target states remain orthogonal and collapse the
superposition.  Thus if the scattering matrix does not depend on ${\bf
P}^{in}$, there can be no interference.

     But if the scattering matrix depends on ${\bf P}^{in}$, there can be
interference in the final momentum distribution of the probe.  We now
illustrate such interference in a simple one-dimensional model
\cite{doron}.  Scattering in this model is elastic and the scattering
matrix is determined---up to an overall coupling constant $\epsilon$---by
(nonrelativistic) energy and momentum conservation. Let $m$ and $M$ denote
the masses of the probe and target, respectively; apart from their
interaction, they are free.  The initial state is the one-dimensional
version of Eqs.\ (\ref{defin}-\ref{instate}).  The final state is the
one-dimensional version of Eq.\ (\ref{finstate}) except that $p^{fin}$ is
determined by $p^{in}$ and $P^{in}$:
\begin{equation}
p^{fin} = {{2m}\over{M+m}} P^{in} - {{M-m}\over{M+m}} p^{in}~~~~.
\label{pfin}
\end{equation}
Thus the scattered part of the final state is
\begin{equation}
\epsilon\int dP^{in}~~ {\tilde \varphi}_\alpha (P^{in}) ~~ \vert
p^{fin}\rangle
\otimes \vert P^{in}+ p^{in} - p^{fin}\rangle ~~~, \label{scatpart}
\end{equation}
and the probability that the probe scatters with a particular momentum
$p^{fin}$ is proportional to $\vert {\tilde \varphi}_\alpha (P_*^{in})
\vert^2$, where $P_*^{in}$ is the value of $P^{in}$ that solves Eq.\
(\ref{pfin}):
\begin{eqnarray}
&{}&{\rm prob}(p^{fin})\cr
&{}&= \epsilon^2 {{M+m}\over{2m}} \left\vert {\tilde
\varphi}_\alpha \left[ {{M+m}\over{2m}} p^{fin} + {{M-m}\over{2m}} p^{in}
\right] \right\vert^2. \label{prob}
\end{eqnarray}
Eq.\ (\ref{prob}) shows that the momentum distribution of the scattered
probe reproduces the momentum distribution of the target, only shifted by
$(M-m) p^{in} /2m$ and scaled by $(M+m)/2m$; and from Eq.\ (\ref{ft}),
$\vert{\tilde \varphi}_\alpha (P)\vert^2$ equals \hbox{$[1+ \cos (Pd/\hbar
+\alpha )] \vert {\tilde \varphi}_L (P)\vert^2$,} where ${\tilde \varphi}_L
(P)$ is broad compared to $\hbar /d$ because $\varphi_L (X)$ is narrow
compared to $d$.  The distribution of ${p}^{fin}$ depends on $\alpha$, as
claimed.

\begin{figure}
\includegraphics[height=6.5cm]{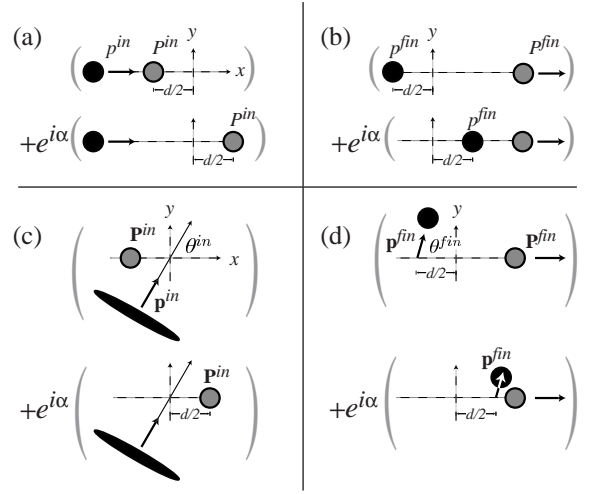}
\caption{(a-b) One-dimensional model: (a) A probe (in black) approaches a
target (in gray) superposed at locations $X=\pm d/2$ with relative phase
$\alpha$.  (b) The probe and target after scattering. (c-d)
Two-dimensional model:  (c) A probe wave packet, with incident angle
$\theta^{in}$ and momentum $p^{in}$, approaches a stationary target
superposed at $X=\pm d/2$.  (d) If the target scatters with momentum
$P^{fin}=Mp^{in}/m\cos \theta^{in}$, the orthogonality of the target
states is transferred to the probe.
\label{fig:schematics}}
\end{figure}

     In any realistic experiment, the incident probe state has a momentum
spread $\Delta p^{in}>0$. To model this spread we fold prob$(p^{fin})$ in
Eq.\ (\ref{prob}) with a distribution $g (p^{in})$:
\begin{equation}
g (p^{in}) = e^{-(p^{in} - \langle p^{in} \rangle )^2 /2(\Delta p^{in})^2}
~~~~.
\label{g}
\end{equation}
Folding prob$(p^{fin})$ with $g(p^{in})$ (i.e. summing
probabilities rather than amplitudes) is allowed because we trace
over the final target state and no two values of $p^{in}$
correspond to the same $p^{fin}$ and $P^{fin}$ (i.e. no two values
of $p^{in}$ interfere in the same final state). To estimate the
visibility of interference fringes, we can approximate $\vert {
\tilde \varphi}_\alpha (P_*^{in}) \vert^2$ by $1+ \cos
(P_*^{in}d/\hbar +\alpha)$ in this convolution.  Then the
probability of $p^{fin}$ is proportional to
\begin{equation}
1 + A \cos \left( \left[
{{M+m}\over {2m}} p^{fin} + {{M-m}\over{2m}} \langle p^{in}
\rangle \right] {d\over \hbar} +\alpha \right)~~~,
\end{equation}
where $A=e^{-d^2(M-m)^2 (\Delta p^{in})^2 /8m^2 \hbar^2}$.  The visibility
of the fringes in the distribution of $p^{fin}$ is, by definition, the
difference between neighboring maxima and minima divided by their sum, so
it equals $A$.  Since the visibility is suppressed exponentially in $d^2
(M-m)^2 (\Delta p^{in})^2 /8m^2\hbar^2$, interference fringes are not
visible for $m\ll M$.  Indeed, $m \ll M$ and Eq.\ (\ref{pfin}) together
imply that the scattering matrix is insensitive to $P^{in}$, as
Cohen-Tannoudji et al. \cite{c} assumed.  But for $m=M$ there is no
suppression of visibility.

     Note that when probe particles of mass $m$ scatter off {\it two}
target particles of mass $M$, visibility is optimal \cite{wp} for $m\ll
M$; here visibility vanishes for $m\ll M$.  This distinction underscores
the novelty of our interference effect.

     We can describe the interference effect more generally as a transfer
of orthogonality.  Initially, the wave function of the target is $\vert
\varphi_\alpha \rangle= [\vert \varphi_L \rangle + e^{i\alpha} \vert
\varphi_R \rangle ]/ \sqrt{2}$, with $\vert \varphi_L\rangle$ and $\vert
\varphi_R\rangle$ orthogonal.  If the initial state of the probe is $\vert
\psi_{in} \rangle$, the overall initial state is $\vert \Psi_{in}\rangle
=\vert \psi_{in} \rangle \otimes [\vert \varphi_L \rangle + e^{i\alpha}
\vert \varphi_R \rangle ]/ \sqrt{2}$ and it evolves according to some
unitary operator $U$ until the probe is detected in a final state $\vert
\psi_{fin} \rangle$.  The probability to detect this final state is tr
$(\rho \vert \psi_{fin} \rangle \langle \psi_{fin} \vert)$ where tr
indicates the trace over the probe and target Hilbert spaces and
\begin{equation}
\rho = U\vert \psi_{in}\rangle \otimes [\vert \varphi_L\rangle +
e^{i\alpha} \vert \varphi_R \rangle ] [\langle \varphi_L\vert +
e^{-i\alpha} \langle \varphi_R \vert ] \otimes \langle \psi_{in}
\vert U^\dagger
\end{equation}
is a density matrix.  Now consider tr$_\varphi (U\vert \psi_{in}\rangle
\otimes \vert \varphi_L \rangle$ $\langle \varphi_R \vert \otimes \langle
\psi_{in} \vert U^\dagger )$, where tr$_\varphi$ indicates the trace over
only the target Hilbert space.  If this latter trace vanishes, then the
probability of any final state $\vert \psi_{fin}\rangle$ of the probe
cannot depend on $\alpha$ and there is no interference.  But tr$_\varphi
(\vert \psi_{in} \rangle \otimes \vert \varphi_L \rangle \langle \varphi_R
\vert \otimes \langle \psi_{in} \vert )$ vanishes because $\vert \varphi_L
\rangle$ and $\vert \varphi_R\rangle$ are orthogonal.  For interference,
then, the superposed states of the target must lose their orthogonality
during the evolution $U$.  The states $U\vert \psi_{in} \rangle \otimes
\vert \varphi_L \rangle$ and $U\vert \psi_{in}\rangle \otimes \vert
\varphi_R \rangle$, however, remain orthogonal as $U$ is unitary. Hence
$U$ must transfer the orthogonality of the target states to other states.
In general, the orthogonal states $U\vert \psi_{in} \rangle \otimes \vert
\phi_L\rangle$ and $U\vert \psi_{in} \rangle \otimes \vert \phi_R \rangle$
are entangled states of the probe and target.  But if they are product
states, then $U$ transfers orthogonality from the target to the probe.
Our one-dimensional model illustrates this transfer.  Fig. 1(a) depicts a
probe approaching a target prepared in the initial state $\vert
\varphi_\alpha \rangle$ of Eq.\ (\ref{defin}), and \hbox{Fig. 1(b)} shows
the particles after scattering.  If $m=M$, the probe and target simply
exchange states, as they do in classical mechanics, so that orthogonality
is transferred from target to probe.  We may regard this exchange as an
interferometric analogue of entanglement swapping.  If $m\ne M$ the probe
and target do not scatter to a product state, but partial transfer of
orthogonality from target to probe still accounts for the partial
visibility at $m\approx M$.

Our general description sheds light also on scattering processes
in which the target is not free, e.g. in a high barrier double
well potential. Here, no transfer of orthogonality is
possible---$\vert \varphi_L \rangle$ and $\vert \varphi_R \rangle$
cannot lose their orthogonality---hence no interference. This
explanation complements the one in Schomerus et al. \cite{s}
ruling out interference on the basis of energy considerations,
when the probe has sufficient energy to excite the antisymmetric
state of the target.


     In our second example, the probe moves in a plane containing the axis
to which the target is confined.  Hence it is scattered by a
``double-slit interferometer" made of a single slit.  The momentum
of the probe has two components, $p_x$ and $p_y$, where the axis
of the target defines the $x$-axis.  Energy and the $x$-component
of momentum are conserved, but not the $y$-component (since the
target is constrained).  We begin with scattering of momentum
states of a probe and target. They scatter at $y=0,~x=X$. It is
helpful to change variables. First, we rescale the position $X$ of
the target to $z\equiv X\sqrt{M/m}$ and correspondingly the
momentum (whether $P^{in}$ or $P^{fin}$) to $p_z \equiv
P\sqrt{m/M}$. With this rescaling, an initial wave function with
momenta $p_x^{in}$, $p_y^{in}$, $P^{in}$ can be written
\begin{equation}
\Psi_{in} ({\bf r},t) = e^{i{\bf p}^{in} {\bf \cdot r}} e^{-i (p^{in})^2
t/2m\hbar}~~~,
\label{psin}
\end{equation}
where ${\bf r} =(x,y,z)$ and ${\bf p}=(p_x, p_y, p_z)$. It
resembles the wave function of a single free particle scattering
on the line defined by $y=0,~x=z\sqrt{m/M}$. Next we rotate
through $\kappa \equiv \arctan \sqrt{m/M}$ in the $xz$-plane,
\begin{eqnarray}
{\bar x} &=& x\cos \kappa -z \sin\kappa ~~~,\cr
{\bar y} &=& y ~~~,\cr
{\bar z} &=& x \sin \kappa +z \cos\kappa~~~,
\end{eqnarray}
and correspondingly
\begin{eqnarray}
{\bar p}_x &=& p_x\cos \kappa -p_z \sin\kappa ~~~,\cr
{\bar p}_y &=& p_y ~~~,\cr
{\bar p}_z &=& p_x \sin \kappa +p_z \cos\kappa~~~,
\end{eqnarray}
so that the scattering line coincides with the ${\bar z}$-axis;
the initial wave function still has the form of Eq.\ (\ref{psin})
but ${\bf {\bar r}}$, ${\bf {\bar p}}^{in}$ replace ${\bf r}$,
${\bf p}^{in}$.  The probe and the target interact at short range,
hence the scattering is cylindrically symmetric; if the initial
momentum is ${\bf {\bar p}}^{in}$ then the probability
distribution of ${\bf{\bar p} }^{fin}$ is
\begin{equation}
{\rm{prob}} ({\bf {\bar p}}^{fin}\vert {\bf {\bar p}}^{in})
= \epsilon^2{{\delta ({\bar p}_z^{fin} -{\bar p}_z^{in} ) \delta ({\bar
p}_\rho^{fin} -{\bar p}_\rho^{in} )} \over{2\pi{\bar
p}_\rho^{fin}}}  ~~,~ \label{wfin}
\end{equation}
where ${\bar p}_\rho \equiv [{{\bar p}_x}^2 +{{\bar p}_y}^2]^{1/2}$.
Transforming Eq.\ (\ref{wfin}) back to the original coordinates, we obtain
\begin{eqnarray}
&{}&{\rm{prob}} (p_x^{fin} , p_y^{fin} , P^{fin} \vert
p_x^{in},p_y^{in},P^{in}) = \nonumber\\
&{}&~\epsilon^2 {{\delta (P^{fin}
+p_x^{fin} -p_x^{in} -P_*^{in}) \delta  (P^{in}-P_*^{in})} \over
{2\pi \tan^2 \kappa \sin\kappa \vert p_x^{fin}-p_x^{in} \vert}}~,~
\label{tofold}
\end{eqnarray}
where $P_*^{in}$ is the value of $P^{in}$ obtained by solving the
two constraints of energy and momentum conservation:
\begin{equation}
P_*^{in} = {1\over 2} \left[ p_x^{fin} - p_x^{in} + {M\over m}
{{(p^{fin})^2 -(p^{in})^2}\over {p^{fin}_x - p^{in}_x}} \right]
~~~~.\label{pin3}
\end{equation}
Now suppose we prepare the target in the state
$\vert{\varphi}_\alpha \rangle$ and the probe in a state $\vert
\psi_{in}\rangle$ with fixed $\theta^{in}$ and a spread $\Delta
p^{in}$ around $p^{in}$.  We obtain the probability distribution
prob$(p_x^{fin}, p_y^{fin})$ for the scattered probe by evolving
the overall state $\vert
\psi_{in}\rangle\otimes\vert{\varphi}_\alpha \rangle$ in time,
projecting onto a final state $\vert p_x^{fin} ,p_y^{fin}\rangle$
of the probe, and tracing $\vert\langle p_x^{fin}, p_y^{fin}\vert
U \vert \psi_{in}\rangle\otimes\vert{\varphi}_\alpha
\rangle\vert^2$ over the final momentum state $\vert P^{fin}
\rangle$ of the target. Expanding $\vert \psi_{in}\rangle$ in
momentum space, we note that since Eq. \ref{pin3} is quadratic in
$p^{in}$ there are at most two values of $p^{in}$ consistent with
the same set $p_x^{fin}$, $p_y^{fin}$ and $P^{fin}$. Hence for
$\Delta p^{in}$ small enough to include only one of the two, we
can obtain prob$(p_x^{fin}, p_y^{fin})$ by summing probabilities,
namely folding Eq.\ (\ref{tofold}) with $\vert {\tilde \varphi
}_\alpha (P^{in}) \vert^2$ and integrating over $P^{fin}$
\cite{feas}:
\begin{equation}
{\rm{prob}} (p_x^{fin}, p_y^{fin}) =\epsilon^2
{{M^{1/2}(M+m)^{1/2}\vert {\tilde {\varphi}}_\alpha
(P^{in}_*)\vert^2}\over {2\pi m\vert p^{fin}_x
-p^{in}_x\vert}}~~~~, \label{arb}
\end{equation}
which we fold with Eq.\ (\ref{g}). Here, as in the first example,
the probe inherits the interference in the initial target wave
function ${\tilde \varphi}_\alpha (P^{in})$.
\begin{figure}
\includegraphics[height=6.5cm]{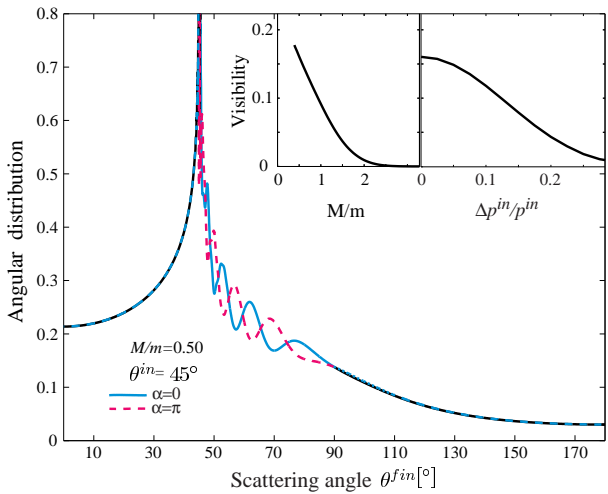}
\caption{Normalized angular distribution of scattered probe
($\Delta p^{in}=0$) incident at $\theta^{in} =45^\circ$, for
relative phases $\alpha =0$ and $\alpha =\pi$ and $M/m =0.5$.
Insets: Visibility as a function of $M/m$, and (for $M/m=0.5$) as
a function of $\Delta p^{in}/p^{in}$, at $\theta^{in} =45^\circ$,
$\theta^{fin}=60^\circ$ \cite{caption}. Here we define visibility
as $[P_\alpha (\theta^{fin}) -P_{\alpha +\pi} (\theta^{fin})
]/[P_\alpha (\theta^{fin}) +P_{\alpha +\pi} (\theta^{fin}) ]$
maximized over $\alpha$, where $P_\alpha (\theta^{fin} )$ is the
probability density for the probe to scatter in the direction
$\theta^{fin}$ from the initial target state $\vert \varphi_\alpha
\rangle$ of Eq.\ (\ref{defin}). \label{fig:results}}
\end{figure}

     If ${\tilde \varphi}_\alpha (P_*^{in})$ changes more rapidly than
the denominator of Eq. (\ref{arb}) as a function of $p^{in}$, we
can estimate the visibility of this interference by approximating
$\vert {\tilde \varphi}_\alpha (P^{in}_*)\vert^2$ by $1+\cos
(P_*^{in}d/\hbar +\alpha )$ as before. We obtain $e^{-d^2
(\partial {P_*^{in}} /\partial p^{in})^2 (\Delta p^{in})^2
/2\hbar^2}$ as the visibility for small $\Delta p^{in}$.  When
$\partial {P^{in}_*} /\partial p^{in}$ vanishes, the visibility is
not suppressed at all, and the spread in the initial state of the
probe is transferred to the final state of the target.  From
$\partial {P^{in}_*}/\partial p^{in}=0$ we obtain the condition
$P_*^{fin} =Mp^{in}/m \cos \theta^{in}$ (where $P_*^{fin} \equiv
P_*^{in} +p_x^{in} -p_x^{fin}$), which we can interpret with the
help of Fig. 1. \hbox{Fig. 1(c)} depicts the probe approaching the
stationary target at an angle $\theta^{in}$, and Fig. 1(d) depicts
the scattering.  The target, initially at $X=d/2$ or at $X=-d/2$,
scatters with momentum $P^{fin}$. If the target was at $X= -d/2$,
it reaches $X=d/2$ after a time $Md/P^{fin}$, while the probe wave
packet requires a time $md\cos \theta^{in}/p^{in}$ to reach
$X=d/2$ if it crosses $X=-d/2$ without scattering.  If these times
coincide, then the scattered target states in the superposition
coincide, and their orthogonality is transferred to the probe. The
condition for this transfer of orthogonality is $P^{fin}
=Mp^{in}/m \cos \theta^{in}$. Since $P_*^{fin} =Mp^{in}/m \cos
\theta^{in}$ is algebraically equivalent to $\partial P_*^{in}
/\partial p^{in} =0$, the condition that visibility not be
suppressed implies transfer of orthogonality, here just as in the
one-dimensional model \cite{pps}.

A full experimental feasibility study will appear elsewhere
\cite{feas}. Let us, however, apply our second example to a
typical experimental setting in which only the final direction of
the probe is measured. We have numerically integrated
$p^{fin}$prob$(p_x^{fin}, p_y^{fin})$ with respect to $p^{fin}$
along lines of constant $\theta^{fin}$. In the numerical
integration, we took $\varphi_L (X)$ and $\varphi_R (X)$ to have
the form $e^{-(X\pm d/2)^2/2w^2}$ with $w=\lambda^{in} /5$, $d=7
\lambda^{in}$ and incident probe wavelength $\lambda^{in}=0.5
\mu$m.  (Double wells with ground state of size 0.1 $\mu$m and
separation 3.5 $\mu$m are achievable with magnetic traps.) While
integrating over $p^{fin}$ tends to average out some of the
interference, \hbox{Fig. 2} shows that the visibility is robust.
The dependence of the scattering on the relative phase $\alpha$ is
very clear. The insets show how visibility depends on the mass
ratio $M/m$ and on $\Delta p^{in}/p^{in}$.  For $M/m>\cos^2
\theta^{in}$ the visibility is suppressed, as we expect since $M/m
>\cos^2 \theta^{in}$ is incompatible with the condition
$P_*^{fin}=Mp^{in}/m \cos \theta^{in}$.

      In summary, we have shown how a distinctive new interferometry can
yield the relative phase of superposed orthogonal location states
of a free target.

\acknowledgments

     Ron Folman would like to sincerely thank Claude Cohen-Tannoudji in
general for years of inspiration and specifically, regarding this work,
for valuable discussions, and Carsten Henkel, Mathias Nest and Peter
Domokos for many fun discussions.  We also thank Doron Cohen for his
insight into the foundations of quantum theory.


\begin{references}
\bibitem{doron}
The original solution is due to D. Cohen (personal communication).  Here we
give it a different interpretation.
\bibitem{wp}
For $m\ll M$ there is no ``which path" information in the target recoil.
\bibitem{caption}For the parameters of Fig. 2 there are no two values of
$p^{in}$ in the range $\Delta p^{in}/p^{in}<0.3$ consistent with
the same $p_x^{fin}$, $p_y^{fin}$, $P^{fin}$.
\bibitem{pps}
Here we can impose the condition of perfect visibility by
postselecting the probe state, i.e. by measuring also $p^{fin}$.
\bibitem{feas}
In a longer calculation, we have generalized Eq.\ (\ref{arb}) to
arbitrary initial probe states, including arbitrary uncertainties
in momentum magnitude and direction.
\end{references}
\end{document}